# The Evenki accounts of the 1908 Tunguska event collected in 1920s – 1930s

Andrei Ol'khovatov

https://orcid.org/0000-0002-6043-9205

Independent researcher
(Retired physicist)

Russia, Moscow
email: olkhov@mail.ru

**Dedicated to the blessed memory of my grandmother ( Tuzlukova Anna Ivanovna ) and my mother ( Ol'khovatova Olga Leonidovna )**

**Abstract.** This paper is a continuation of a series of works, devoted to various aspects of the 1908 Tunguska event. It is devoted to the Evenki accounts of the 1908 Tunguska event collected in 1920s – 1930s. It is important to research accounts of Evenki who were rather close to the epicenter. The Evenki accounts are important also, because Evenki are natural hunters and pathfinders – their lives depend on their memory and vision. Most of the reviewed in this work accounts were collected at the Evenki conference, when telling a lie was considered to be a serious misconduct. These Evenki accounts are compared with other Tunguska accounts. Also weather conditions associated with the Tunguska event are considered. Some manifestations of the Tunguska event are discussed.

## 1. Introduction

This paper is a continuation of a series of works in English, devoted to various aspects of the 1908 Tunguska event [Ol'khovatov, 2003; 2020a; 2020b; 2021; 2022; 2023]. It is devoted to the Evenki accounts of the 1908 Tunguska event collected in 1920s – 1930s. Also some results obtained in later times, especially by members of KSE (the Complex Amateur Expedition) are considered. The Evenki accounts are important, because Evenki are natural hunters and pathfinders – their lives depend on their memory and sight. Moreover most of the examined in this paper accounts were collected at the Evenki conference in 1926, when telling a lie was considered to be a serious misconduct.

Krinov wrote in his book [Krinov, 1949] about the Evenki accounts (translated by A.O.):

"Nevertheless, looking ahead, we should note that many of the stories, as it was later established during the expeditions, turned out to be remarkably accurate."

KSE practically stopped sending expeditions to collect eyewitness accounts in the late 1970s, deciding that the reliability of new accounts obtained through third parties would not pay off the efforts to collect them. Indeed, for example, the article [Kavková, et al, 2022] says (TE is the Tunguska event): "According to local nomads (Evenkis), Suzdalevo Lake did not exist before the TE...", but according to the result of the research: "Therefore, Suzdalevo Lake was evidently formed before the TE".

Let's compare with the time when the eyewitnesses were still alive. In 1958, an expedition of the USSR Academy of Sciences surveyed a small lake, which, according to the Evenks, did not exist before the Tunguska event. According to the results of the survey, it turned out that the trees that fell during the Tunguska event formed a blockage that blocked the riverbed. Subsequently, the blockage was covered with silt and various plant debris, and a lake formed above it. Later, the dam was washed away and the lake disappeared. That's why the early accounts are very important.

In the morning of June 30, 1908, thunderous sounds were heard by population north and northwest of Lake Baikal in Central Siberia, In some places also the ground trembled. Reports of a flying glowing body came from various points of the region. Soon a newspaper story appeared about a fall of a large meteorite near the town of Kansk, but then it was recognized as wrong. Some years later a large forestfall of radial character was attributed to a probable result of the event. The forestfall is named after the first scientist who researched it – Leonid Kulik.

On Fig.1 there is a map of the region (based on English edition of [Fesenkov, 1966] with some additions). Red arrows drawn by the author (A.O.) 1, 2, 3 are proposed (by various researchers) trajectories of the alleged Tunguska spacebody.

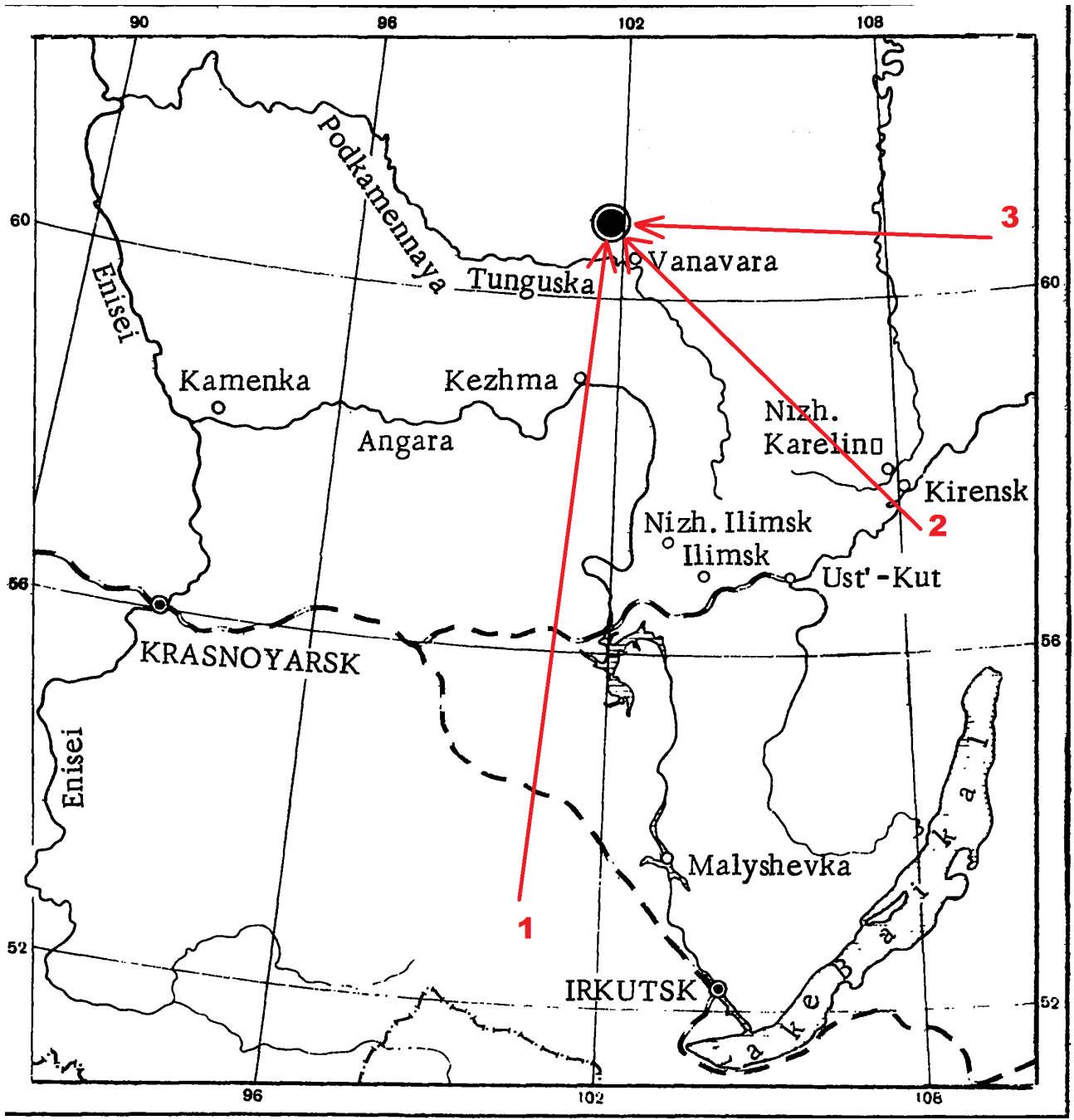

**Fig.1**

The epicenter's (center's) position of the Kulikovskii forestfall is about 60°53' N, 101°54' E (it depends on a method of calculation). In this paper the distance will be often measured from the epicenter of the Kulikovskii forestfall (and azimuth is measured from north clockwise). The data is from [Vasil'ev, et al, 1981] mainly.

The Evenki accounts (in this paper) are limited pre-WW2 time in general.

## 2. The accounts

The chain of events which led Kulik to the place north of Vanavara can be seen from his 1927 article, which was rather well translated in English in 1935 [Wiens and La Paz, 1935]. Here are some citations from the translation regarding Evenki (Tunguse) account from the winter of 1908/1909:

"Thus, a member of the management of the Consumers Association of Kezhma on the Angara, I. K. Vologzhin, reported the following to the author in the city of Krasnoyarsk on the 21st of November, 1921: "In the year 1908, on June 17, I was about 20 versts above the village Kezhma (along the river Angara) in the market village Chirida. In the morning we examined our fishing nets. It was a bright morning. There was not a cloud. With me was an old man. We heard—a clap, then—another one, and gradually the thunder rumblings began to diminish toward the north. In the winter the inhabitants of the village Kezhma, who traded around the Podkamennaya Tunguska (Khatanga), reported that the Tunguse Ivan Ilyich Ilyushenok related that at the time of this occurrence, in the locality where they were wandering between the Podkamennaya Tunguska and the Nishnaya [lower] Tunguska, a strip of forest was uprooted by the pressure of the air and several of his reindeer were killed. People suppose that the place of fall is in the region between the Podkamennaya Tunguska and the Nishnaya Tunguska." "

However unfortunately Kulik omitted one sentence from inside the account, and did not mark this with a relevant sign. Before the words "In the winter…" Vologzhin wrote (translated by A.O.) [Vasil'ev, et al, 1981]: "I don't remember what time it was, but the sun hadn't risen yet, it was just dawn." Unfortunately it is not the single case when accounts were edited in similar ways without any notice. In [Ol'khovatov, 2008, 2020a] several examples are given, when accounts were even ignored, probably because they contradicted ideas of some Tunguska researchers.

Let's continue with translation from [Wiens and La Paz, 1935].

"I. I. Pokrovsky (city of Yennisseisk [Yeniseisk], former financial inspector) corroborates the rumor about the breaking down of the forest by a wind blast; in a letter to the author under date of April 21, 1922, he writes : "According to the testimony of the Tunguses, the falling of the meteorite was accompanied by an unusual atmospheric perturbation which caused a terrible destruction of the woods over a large area." "

I. V. Kolmakov (co-operator in the village Panovskoye on the Angara) wrote to

Kulik on February 10, 1922 the following [Wiens and La Paz, 1935]:

"

"<...> Moreover, I, personally, talked with a Tunguse about this [event]. The latter related the following: 'Along that same river Tunguska in the region of the junction with the Chamba over a distance of 350 versts from Panovskoye, during the time of this thunder, about a thousand of the Tunguses' reindeer were killed and the remainder of the reindeer badly injured, and also the natives themselves suffered from the heavy shake [concussion], and, moreover, in a region of radius approximately 70 versts, all the forest was destroyed and right there by the shock a spring was opened in the earth which disappeared after several days, but the place of outflow of the water was not examined by the natives. All that which is contained herein I affirm under oath."
"

Here is a fragment of a letter dated January 19, 1924, by the engineer V. P. Gundobin, who lived for two years near the Podkamennaya Tunguska, close to the region of the event [Wiens and La Paz, 1935]:

"

"The second time I heard about this same phenomenon was later in April, 1921. It was from Tunguses who said that they heard a heavy roaring thunder which frightened them terribly; they, also, spoke of the destruction of reindeer. <…>  From the same I. V. Kokorin, I learned that the sounds were still greater on the Podkamennaya Tunguska. Having come to this river, Kokorin met two Tunguse brothers (one of them was Ivan Ilyich Onkoul). These related that something flew down from the sky and broke down the forest, after which a fire started in which twenty of their 'reindeer' were consumed. It is necessary to point out that the expression 'reindeer' is used also as a synonym for that amount of flour which one reindeer can carry [as a permanent load] on a trip; so that the expression 'twenty reindeer' may be understood in either the literal or the figurative sense, that is, that a shed containing from 80 to 100 puds burned down. The wandering region of these Tunguses who live on the Kezhma is situated in the district of the river Ognia, a tributary of the river Vanovara. This phenomenon occurred in the administration of Prince Dushinchi whom I saw and who affirmed that actually at the time previously referred to, there was a fire on the river Ognia, that a mountain was shattered to pieces there, and that this region is considered accursed by the natives. From inquiries made of other Tunguses, the idea presented itself that along the river Taimur (district Vanovara), a region two or three versts in extent was strewn with rocks [at

the time of the fall], at which time several families and many reindeer perished."
"

One pud is about 16.4 kg. The place of reported destruction near the river Ognia is about ~55-60 km ( at azimuth about 128 degrees ) from the epicenter. Remarkably that according to Kulik [Kulik, 1937]:

"The investigation brought to light a continuous, eccentric, radial "windfall" in a huge area of radius about 30 km., which extended on some of the hills to a distance of 60 km. and even farther. Observers relate that individual trees were thrown down on the hills in the neighborhood even of Vanovara."

Krinov also wrote [Krinov, 1949] that the first traces of the taiga's damage were near Vanovara (translated by A. Ol'khovatov, factory here means "trading post" ):

"If you get acquainted with the attached map, you can see further that the first traces of the action of the explosive wave on the periphery of the region are noticed almost at the Vanovara factory itself, where ( especially along the riverbed of the Chamba River, starting from its mouth ) trees with broken tops growing on the shores are observed. Such trees sometimes are met alone, sometimes in groups with several trunks located nearby."

The statement about a forest-fire (near the river Ognia) at such distance from the epicenter is very remarkable. Moreover the river Taimur (Taymura) is about 300 km NE of the epicenter (but a confusion with topographic names can't be ruled out). So early Kulik thought that the meteorite fell in the area of the river Ognia - the left upper tributary of the Vanovara (Vanovarka) river.

In 1924 Kulik received a letter dated February 2, 1924, from the geologist and co-worker in the Krasnoyarsk Museum, A. N. Sobolev. Here is what Kulik wrote about it [Wiens and La Paz, 1935]:

"His account gives a vivid picture of this mighty phenomenon and completely agrees with the data of I. M. Suslov: "A certain N. N. Kartashev working in the summer on the Podkamennaya Tunguska reports: 'According to the story of a Tunguse Ilya Potapovich [no surname], who lives on the river Tetera in the upper regions of the Podkamennaya Tunguska, long ago (about 15 years), there lived on the river Chamba his brother [who is now an old Tunguse and who hardly speaks Russian, and who now lives on the Tetera with Ilya Potapovich and whom N. N. Kartashev saw]. Ilya Potapovich reported that there [on the river Chamba], once upon a time, occurred some sort of terrible explosion with noise and wind. The power of the explosion was so great that, on the river, for many versts along both sides, the forest was broken down in one direction. The reindeer skin tent of his brother was thrown down, the top of the tent was carried away by the wind, the brother was deafened, and the reindeer were thrown into a panic. On regaining composure, he was not able to round up but very few of the reindeer. All this so worked upon him that he was sick for a long time. In the broken-down forest at one place a hole was formed out of which flowed a little brook into the river Chamba. Through this locality [i.e., where the woods were broken down] formerly went a Tunguse road; now it is abandoned, because it turned out to be all obstructed [by the broken down trees] and impassable, and, besides, for the reason that it caused horror in the Tunguses. <…>"."

Kulik was aware about the Evenki accounts collected by S.V. Obruchev in 1924 [Obruchev, 1925]. Obruchev marked that Evenki rejected the fact of the meteorite fall, but were ready to show the area of the forestfall near the Chamba river. Obruchev wrote (translated from Russian with remarks in {...} by A. Ol'khovatov) [Obruchev,1925]:

"In the summer of 1924, I was sent by the Geological Committee for research of the r. {river - A.O.} Podkamennaya Tunguska. During the work, I assumed to attend the place of the meteorite fall. Unfortunately, I failed.
<...>
Tungus, about whom it is definitely known that he saw a hole punched by a meteorite during a fall, said that he had never been to this place, but tells from the words of others wandering there. The latter denied in general that the meteorite had fallen, and agreed to show only the area of the fallen forest. The latter, according to them, reaches 680 sq. km; ... <...>

Tungus Ilya Potapovich, who lived in the Teterya trading post, informs that there is a pit in the place where the stone fell, a stream comes out of it to Chamba. There is a lake nearby, but it existed before the fall a meteorite. His brother was standing just at the time of the meteorite fall in this area, his tent flew up into the air, "like a bird", the deer were partly killed by falling trees, some fled, and he himself lost his tongue from fright for several years.

The rumble of the meteorite was heard both in the Teterya and Vanovara factories {trading posts – A.O.}, on the Podkamennaya Tunguska and on the Angara river in all visited by me in 1924 villages from s. {settlement – A.O.} Dvorets to s. Panovskoe. The rumble was heard in the morning (according to other accounts, in lunch {"obed" - "dinner" in Russian – A.O.} - i.e. about 10 hours). The {window's – A.O.} glass was trembling in the north side, the objects fell from the shelves, in one case the horse on which they drove, fell. In the Teterya factory {trading post – A.O.} fiery pillars were seen in the north."

Let me make some comments. The fiery pillars seen from Teterya (Tetera) are remarkable (the Tetera factory is about 83 km to SSE from the epicenter), as well as 2 times given are intriguing (about the problem of the times see also [Ol'khovatov, 2022, 2023]). A possible explanation is presented in this paper below.

The reported size of the fallen tree area is close to the size of the 'total uprooting' area – see [Ol'khovatov, 2021], and should not be confused with the area (about ~2150 sq. km) where even minor traces of the forest-fall (assigned to the event) were discovered.

Now about the "lake nearby" – probably it is the Cheko lake. It is interesting to compare with an account by Evenk Andrei Dzhenkoul which is written in the Nikolai Vasil'ev's diary of 1974 (from tunguska.tsc.ru). According to the account, the Andrei's father was at the Cheko lake, they were thrown into the air, tents were thrown away, and that was all (the end). But at first the sky turned red, and then an awful blow occurred. It is useful also to compare with words by E. Krinov [Krinov, 1949] translated by A. Ol'khovatov (Krinov called the Cheko lake as Lebedinoe):

"From the tops of the hills located to the north of the basin (or to the west of the Mount Farrington), the author could clearly see the perfectly preserved powerful taiga in the area beyond the Kimchu River, near the Lebedinoe lake, where, apparently, there is a hill that the Kimchu {river –A.O.} surrounds. The fully preserved forest on this elevation was clearly visible.".

It clearly demonstrates that in the direction of the Cheko lake (the lake is just about 8-9 km from the epicenter), the effect of the explosion was relatively weak (see also comments by J. Anfinogenov, and L. Budaeva below).

Also Kulik was informed about Evenki accounts collected by I.M. Suslov in 1926 before they were published in 1927 [Suslov, 1927]. Being the Chairman of the Krasnoyarsk Committee for Assistance to Northern Peoples, Suslov had to visit the area near the forestfall to prepare the election of the first agencies of the Soviet power in Evenkiya – the Tribal Councils and Tribal Courts. It was for the organization and preparation of Munniak/Suglan (a big meeting of Evenki).

In his article Suslov also presented a map compiled him from the accounts. It is interesting that authors of [Anfinogenov, and Budaeva, 1998a] evaluate the accounts in the following way, translated by A.O. (see also [Ol'khovatov, 2021] for

details):

"Thus, after interpretation of aerial photographs of 1949 it was possible to establish the configuration of the clear-cut (total uprooting) zone. The fact of a break in the clear-cut zone in the western sector was established (Fig. 2, 3). This fact was confirmed during field work in 1967. An explanation was found for the previously considered incomprehensible break in the forest clearance boundary on the western side of the scheme drawn by Suslov based on testimonies of Evenks in the early 1920s. It is worth noting the accuracy and attentiveness of the Evenki observations. Obruchev, Kulik, and Florensky had solid boundaries on the western side of the fallout zone, although Florensky had dotted lines."

In the mid-1960s Suslov presented more detailed data on accounts of 1926, which were published in 1967. Here the data will be considered, basing on English translation of the 1967-version from [Suslov, 2006]. The author of this paper has re-worked the translated text so that, on the one hand, it is as close as possible to the original, and on the other, to be as clear as possible and avoid ambiguous interpretation. In addition, the found typos have been removed and small omissions have been restored.

Let's start with the Akulina account (please pay attention that 'chum' here means a portable dwelling in the form of a conical tent covered with hides, bark, felt, etc.).

"Not far from the faktoriya (trading station) of Vanavara I saw the chum of Ilya Potapovich (Luchetkan), an Evenk, in whose family lived also Akulina (the widow of his brother Ivan). In June 1908 their chum stood at the mouth of the Dilyushmo river in the place of its confluence into the Khushma river. Akulina recounted about this event in following words:
– "We were three in our chum – I with my husband Ivan, and the old man named Vasiliy (son of Okhchen). Suddenly, somebody pushed our chum violently. I was frightened, gave a cry, woke Ivan and we began to get out of our sleeping-bag. Now we saw Vasiliy getting out as well. Hardly had I and Ivan got out and stood up when somebody pushed violently our chum once again and we fell to the ground. Old Vasiliy dropped on us as well, as if somebody had flung him. There was a noise all around us, somebody thundered and banged at the elliun (the suede cover of the chum). Suddenly it became very light, a bright sun shone at us, a strong wind blew at us. Then it was as if somebody was shooting, like the ice breaks in the winter on the Katanga river, and immediately after that the Uchir dancer swooped down, seized the elliun, turned it, twirled it, and carried it off – somewhere. Only the diukcha (the chum's framework, consisting of 30 poles) has remained at its place. I was frightened to death and became bucho (lost consciousness), but now see – the uchir (whirlwind) is dancing. I gave a cry and came back to life (regained consciousness).

The uchir hurled the diukcha on me and hurt my leg with a pole. Now I got out from under the poles and began to cry: the small chest/box with plates and dishes is thrown away from the chum and is lying at a distance, it is open and many cups have been broken. I am looking at our forest and cannot see it. Many trees stand without branches, without leaves. Plenty of trees lie on the ground. Dry tree-trunks, branches, deer moss are burning on the ground. Now I see some clothes burning, I came and see – this is our blanket from hare-skin and our fur bag, in which I and Ivan slept.

I went to look for Ivan and the old man. Now I see something hanging on a twig of a naked larch. I approached, stretched a stick and took it down. This was our peltry that (earlier) had been suspended (tied) on the chum's poles. Fox pelts was scorched, ermine became yellowish, dirty and sooty. Many squirrel's skins have broken into wrinkles and got too dry.

Now I took our peltry and went, crying, to search for my men. And the dry wood was still burning on the ground, deer moss was burning, the air was filled with smoke.

Suddenly I hear somebody moaning softly. I ran to the voice and saw Ivan. He was lying on the ground between the branches of a big tree. His arm has broken over a log, the bone showed through the shirt and protruded on the outside, with blood dried on it. Now I fell down and became bucho again. But soon I returned to life again. Ivan has 'awaken' and began to moan louder and to cry.

The uchir threw Ivan near. If ten chums are placed in sequence then he fell down behind the last chum, quite near to the place where I took off the peltry from the twig [A diameter of a chum is about four meters; therefore, Ivan was thrown away for about 40 meters. – a comment by Suslov].

Well, Ivan put his healthy arm round my neck. I picked him up and we moved towards our chum at the Dilyushma, where there were in a labaz (storehouse) two elk skins, a bag of flour, and (fishing) nets. The chum stood at a bank of the Dilyushma river, the labaz was not far from it, to the sunset. Suddenly we seem to hear as somebody is shouting. Now we happened to see our Vasiliy. He got under the roots of a fallen larch and hid there. I got tired, handed Ivan over to the old man, and I got to carry the burned peltry alone.

Now walking became even more difficult: there were so many fallen trees around. Suddenly we saw on the ground logs and elk skins under them. The hair on the skins had burnt, the skins themselves broken into wrinkles and also burnt. Instead of nets, we saw a heap of small stones – sinkers. The nets made from the horsehair had completely burnt. The logs had also burnt out, turning into firebrands. Instead of the bag of flour – just a black stone. I have poked it with a stick and the stone-coal broke. Inside it I found some flour and rolled the flour up in the Vasiliy's shirt. That's how our labaz has perished. After a short rest, we went to look for our chum.

And now here it is, the place where our chum had stood. The poles lie on the ground, with a large fallen larch on them, the latter having been much burnt. I have cut it with my axe and dragged it aside. Under the larch we have found our copper cauldron in which there was a lot of yesterday's meat.

A light summer night has fallen. The fire diminished. It became cold instead of heat. We decided to go to the Katanga. When we reached the Chamba river, we were already very weak. And we saw around us a miracle, a terrible miracle. The forest was not our. I have never seen such a forest in my life. It was so unfamiliar. We had here a dense forest, a dark forest, an old forest. And now there was in many places no forest at all. On the mountains all the trees were lying down and it was light; one could see far away. And it was impossible to go under the mountains, through the bogs: some trees were standing there, others were down, still others were bent, and some trees had fallen one upon another. Many trees were burnt, dry wood and moss were still burning and smoking. Having come to the Katanga we met with Luchetkan."

I have presented the Akulina's story here in abbreviated form: I have omitted some details having no direct relation to the Tunguska catastrophe."

In his 1927 article Suslov also mentioned that two weeks later at the trading station of Strelka, he met Vasiliy (son of Okhchen), who was living with Ivan and Akulina during the event. He confirmed the Akulina's account, and added that he woke up at the moment when the tent was torn away, and he was thrown aside by a powerful push, but did not lose consciousness. An incredibly strong prolonged rumble was heard, and the ground shook. Burning trees were falling, and everything round about was shrouded in smoke and fog. Soon the rumble died away, the wind ceased, but the forest went on burning. All three set off in search of the reindeer which, at the time of the catastrophe, had rushed away. Many of them did not come back and could not be traced.

Suslov noted that at the time of fall Akulina's tent stood at  the  mouth of the Dilyushmo river near the place of its confluence into the Khushma river, but the River Dilyushmo does not exist on modern maps. There are two rivers "similar in names" - Nizhnyaya Dulyushma and Verkhnyaya Dulyushma, which flow into the Chamba river (in the area not far from the mouth of the Khushma). The problem is that in those years this area was practically unexplored, there were no detailed maps, so Suslov probably used local names. Here are just several examples how poorly this area was investigated.

L. Kulik in the article "Where the Tunguska meteorite fell" (published in "Nauka i Tekhnika" N 39 in 1927) wrote (translated by A.O.):

"... meanwhile, I had to work in such a region where all the elements of our geographical maps are so conditional that it would be more correct, perhaps, to designate this entire region for the time being as a solid white field. Suffice it to say, for example, that for the entire Chunya River and the upper reaches of the Podkamennaya Tunguska there is only one astronomical point (the Taimba trading post), and tails of these rivers (almost 1000-verst {maps - A.O.}, which have already been captured by the route survey) hang very arbitrarily in space; therefore, the question of one Krasnoyarsk cartographer addressed to me (as to an expert, who has visited the areas), is understandable: what should he do when compiling the 20-verst and  80-verst map, at least with the same river Chunya - "stretch it a little or compress it with/as an accordion.""

B. Bronskiy wrote in his diary on July 28, 1960 (translated by A.O.):

"So for example, Ogne (a tributary of the Chamba on the hundred thousandth {1:100000 scale - A.O.} map) is actually called by the Evenks Elyumba, and what is marked on the map as Elyumba, the Evenks call Ogne."

J. Anfinogenov, and L. Budaeva wrote  [Anfinogenov, and Budaeva, 1998b] (translated by A.O.):

"We especially note the presence of toponymic difficulties. So, on some pre-war maps, the Chamba River is marked as the "Kh" river, as the Khusma River, as the Vanavarka River. It is possible that when working with eyewitnesses there was a combination of names."

Anyway according to the map in the 1927 Suslov's article the Akulina tent/chum was near the mouth of the Khushma river. On Fig.2 (based on a sketch by W. Fast) it is marked as the number "1" in red by the author of this paper. The epicenter is indicated by a red square. The Akulina's chum was positioned at distance about 32 km and azimuth about 129° from the epicenter.

**Fig.2**

The next important account obtained by Suslov  was from  brothers Chuchancha  and  Chekaren [Suslov, 2006] (slightly adapted and added):

"From  the  faktoriya  of  Vanavara  I  left  for Strelka-Chunya. There I met with the brothers Chuchancha  and  Chekaren  of  the  Shanyagir.

At  the  moment  of  the  catastrophe  of  1908, their chum stood near the chum of their father in  the  middle  reaches  of  the  Avarkitta river. Both  brothers  proved  to  be  interesting  and intelligent  interlocutors  and narrators.  They spoke  Russian well enough  in  the  Angara-Tungus dialect/slang.  Chuchancha  has  essentially repeated Akulina's description of events, but I have  asked  him  to  recall  how  many  thunder blows – "agdyllian" – occurred  and  how strong  they  were.  In  the Chuchancha's words, he counted five blows.

"Our  chum  stood  then  on  the  bank  of  the Avarkitta.  Before sunrise  I  and  Chekaren came from  the Dilyushma river  where  we stayed with Ivan  and  Akulina.  We fell asleep quickly.  Suddenly  both  of us awoke.  Somebody  gave  us  a  push.  Then  we heard  a  whistle  and felt  a  strong  wind. Chekaren  even  exclaimed: 'Do  you  hear? There are so many goldeneyes or mergansers flying.'  We  were  still  inside  our chum  and could not see what was happening in the forest.  Suddenly  I  got another  push  from  somebody – so  strong  that  I  knocked  my  head against the chum's pole and then fell onto the hot  coals  in  the  hearth.  I got  frightened. Chekaren  had  also  got  frightened and snatched at the pole. We began to call our father,  mother  and  brother,  but  nobody replied. Noise was heard from the outside of the chum. Trees could be heard falling. Chekaren and me, we got out from our sleeping bags and were going to go out of the chum,  but  suddenly  there  was  a  very  great clap  of  thunder.  This  was  a  first  blow.  The  earth  started to jerk,  and swing,  a  strong  wind  hit  our  chum and  threw  it  down. I was firmly pinned down by the poles, but my head was not covered, because the ellun was lifted up. And now I saw  some  terrible miracle:  trees  were  falling down,  their  pine-needles  burning.  Dry wood on the ground and  deer moss were burning  as  well.  Smoke  is  all  around.  The  eyes  ache,  it's hot,  very  hot,  one  could  burn  out.

Suddenly  there  appeared  above  a  mountain, where the trees had already fallen down, a bright light (appeared) like a second sun appeared. The Russians would have said: it has unexpectedly flashed. It hurt my eyes, and I even closed them. It was like what the Russians call 'molniya' (lightning). And at once an agdyllian, strong thunder, crashed. This was a second blow. The morning was sunny, no clouds, our sun shone bright as always, and now a second sun!

With an effort I and Chekaren got out from under the chum poles and elliun. After that we saw like a flash appears again and a strong thunder heard again overhead, although in another place. This was a third blow. The wind came at us, knocked us off our feet, hit us on a fallen tree.

We were looking at falling trees, we saw as their tops break, we watched the forest fire. Suddenly Chekaren cried out: 'Look up!' and showed with his hand. I looked there, and I saw a lightning, it flashed and struck again, made an agdyllian. But its sound was not so loud as before. This fourth blow was like a usual thunder.

Now I remember well that there was a fifth blow, but rather weak and far away."

- "But at which side did you hear this last, fifth thunder?" - I asked the brothers.

- "It was where the sun is sleeping during the night, where the Taymura river is" - replied Chuchancha.

I attempted to make a quantitative estimate of the lapse of time between the first and the second blows. It could be done only through comparison with a time interval usual and well-understood by the Evenk hunters.

I decided to use the effect of echo. Several days earlier, I was making a route survey of the environs of Strelka on the Chunya river. One of my routes went along the Northern Chunku-kan river, where there was a cliff/rock not far from the Strelka. I brought with me the brothers Shaniagir to the cliff-face of the 'Syrka' cliff, from which one could see both the place of confluence of the two rivers Chunku-kan into one river - Chunya, and the cliff/rock. The distance between the rocks is 1020 meters.

- "Look there," - I told the brothers. - "Stepan Ivanovich, please shoot from both barrels towards that cliff/rock at Chunku-kan. First we will hear the report, which will be as the first clap of thunder, and then we will hear its echo, which will be the second clap of thunder. Note how long or short will be the interval between the shots and the echo."

We went down to the water. Chuchancha and Chekaren shot several times, in turns, and I noted the moments of shooting. Naturally enough, the echo was heard 6 seconds after each shot. Both brothers stated that just the same time interval was between the first and the second claps of thunder. "

The report of the brothers (that their chums were situated near the River Avarkitta)

was confirmed by E. Krinov when he discovered the remains of these chums in 1929 [Krinov, 1949]. Unfortunately now it is not possible to precisely position the chums. Approximately their chums were 37 km at azimuth ~160° from the epicenter. On Fig.2 it is marked as the number "2" in red by the author of this paper.

Let's continue with the accounts collected by Suslov:

"Before the closing of Munniak I appealed to all its participants, asking them to verify the facts reported by Akulina, Vasiliy (the old man), and the brothers Chuchancha and Chekaren, hoping also to obtain additional data. It was not easy task. I knew that the Evenks attributed the catastrophe of 1908 to shaman Magankana's revenge, carried out by a flock of iron birds (agdy). Trying to refute this idea then was utterly useless, especially since the Evenks believed the old man Vasiliy of the Shaniagir kin (who also attended the Munniak) "had seen these birds himself" and "saw how they rattled, made noise, hit loudly".

Therefore, I had to find a roundabout way to get the people to talk about the subjects that were forbidden by their religious "taboos." I addressed the delegates asking them to tell in some detail who of them had suffered through the catastrophe; where the zone of tree leveling began and where it ended, as well as whether somebody had seen holes in the ground that were absent before the fall of a meteorite.

An excited conversation started among the Evenks who were sitting in the glade. There even arose a heated debate. Then the old man named Ulkigo (a son of Lurbuman) of the Shaniagir kin, began to speak. It was believed Ulkigo was 80 years old.

"The chum of my father Lurbuman stood on the bank of the Chamba river, not far from its mouth. There lived in this chum my father, I with my wife, and our four children. One day, early in the morning, dogs suddenly started howling and the children began to cry. My wife, I and the old man awoke and saw a miracle. We began to listen and heard somebody begin to knock on the ground below us and swing the chum. I jumped out from the sleeping-bag and started to dress myself, but now somebody pushed the ground violently. I fell and shouted, the children shouted too and jumped out from sleeping-bags. Shortly before that, somebody was strongly shooting from guns. Well, the old Lurbuman said a rock/cliff had fallen near the Churgim stream. Suddenly again – as if somebody knocked at the ground very strongly, a copper kettle fell down from a pole in our chum and somebody made the Angara thunder [The Evenks call thunder "agdy", whereas the Angara inhabitants call thunder "grom". So it was a blow of thunder, literally a blow. – comment by Suslov]. I dressed up soon afterwards and ran out from the chum. The morning was sunny and cloudless. It was hot. I began to look upwards at the Lakura mountain. Suddenly something flashed brightly in the sky and the thunder crashed. I got frightened and fell down. Now I looked and saw the wind felling the trees and the fire scorching the dry wood on the ground. I heard a noise somewhere. Now I jumped to my feet and saw two elks with their calf and two deer run to the Katanga. I got a fright and went back to my chum. At this moment the Uchir swooped, took the elliun and threw it to the river. Only the diukcha remained. Near it there were sitting on a leveled tree my father, my wife, and the children.

We are looking at that direction where the sun sleeps (that is, at the north). Some miracle is occurring there; somebody is knocking again. In the direction of the Kimchu river one can see large smoke, the taiga is burning, strong heat is felt from there. And suddenly, somewhere far away, at the Chunku-kan river, a strong thunder crashed again and smoke appeared there.

I went to look over that location, from which the wild animals were running and the heat was coming. There I saw a terrible miracle. The whole taiga has leveled, many trees on the ground were burning, dry grass and dry twigs were burning, all the leaves in the forest got dry. It was very hot with much smoke. The smoke stung the eyes. It was completely impossible to look around. I was very scared and ran back to the Chamba, to our chum. When I told my father about all this, he got a fright and died. The same day, we buried him according to our Tungus belief." "

The Ulkigo's chum was positioned about 50 - 56 km from the epicenter at azimuth about 176°. On Fig.2 it is marked as the number "3" in red by the author of this paper. Please note that the positions of all eyewitnesses are shown approximately. Let's continue with the accounts:

"Many other Evenks, who participated in the Munniak, told something like the following:
"The agdy birds struck many times, strongly struck. The agdyllian [Thunder. – comment by Suslov] knocked, the Angara thunder made blows, the tan'ga struck five blows. Paktyrun – as if shot. It was, however, Uchir that was burning the forest and leveling the trees. It took away Akulina's elliun to somewhere and threw the elliun from the Ulkigo's chum to the river. Odyn [Odyn – storm, squall – is represented in the Evenk's demonology as a hydra-headed monster with enormous mouths and without any eyes or tail - comment by Suslov] was leveling the diukcha, it spoilt Akulina's squirrels. It was ruining, burning, leveling labazes. The agdys were ruining deer, dogs, and harmed some men – three people have died:  Lurbuman,  Ivan  Machakugyr (Luchetkan's brother) - has hurt his arm and died, Uyban (the  shaman)  has  become  bucho at once and died on the Lakura."

Some personal statements were then made as well.

Andrey Onkoul: "I was looking for deer between the Lakura river and the Kimchu river. There I saw a pit and a dry river running from it.  This  is  on the Lakura  mountain. Before the 'sorrow/grief' there was no pit at this place, neither the 'dry river' (furrow)."

Molok  Kurkagyr: "Half  a  day  of  niulgui (that is, a half of a day's march in summer on pack deer) from the Chunku-kan river to mid-day  (that is, to the south)  the taiga was also leveled strongly; a large pit was made. The trees were there lying, their tops towards Erbogachen.  Earlier,  before the  'sorrow/grief'  there was no pit there; the forest was dense, with a lot of squirrels."

Luchetkan:  "On the mountain, on the Lakura range, near that place from which a stream  and  then  the  Markitta  river are flowing,  Akulina also  saw  a  'dry  river' (furrow). At the end of this river there was a big  pit filled  up  with  soil. There were many leveled trees there."

Such were the main  testimonies  obtained from  the Evenk witnesses, and  some  of  who experienced  the  terror  of  the  catastrophe of 1908.

<…>

In  1926  and  somewhat  later,  both L. A. Kulik and I analyzed more than once the eyewitnesses' reports  about  "several  claps  of thunder" and  "flashes  that occurred  somewhere high," admitting  that  the meteorite could probably explode in the atmosphere and several lumps could fall down. L. A. Kulik was so  sure  his  hypothesis  was  correct  that in 1927 he mistook the  small  circular  lakes  located  at  the  Great marsh  (so  named  by L. A. Kulik)  inside  the  cirque  Merrill, for  craters formed  by  large  meteorite  fragments.  It was only in the 1930s that he agreed that both the  unevenness  of  the  marshes  inside  the  cirque Merrill and these "craters" were due to usual  thermokarst  processes, typical  for  the subarctic zone of permafrost.

Probably, because of this, the reports of the witnesses about several explosions do not attract any longer the attention of researchers.

At present, when a large team of researchers is examining the problem of the Tunguska catastrophe of 1908 in great depth, it seems to me rather interesting to publish a more detailed exposition of eyewitnesses reports than had been done in the article of 1927."

Unfortunately the Kulik's expeditions in the late 1920s – 1930s did not add much to the description of the event. Here is a fragment of a 'second-hand' account by K.A. Kokorin written down by E.L. Krinov in 1930 [Vasil'ev, et al, 1981] (translated

by A.O.):

> "Tungus S.I. Ankov ... with three their brothers ... in the same year when the meteorite fell, in winter (in January) they came to the trading post Panolik and he, Kokorin, was told that 80 tursuks/bags of flour and warm winter clothes, which were in the warehouses near the Lakura ridges, burned down from the fall of the meteorite. Deer also died there for a part. When they came to the warehouses (after the fall), they saw on a flat place a "rupture of the earth" in the form of a large ditch without water, in which they found all sorts of pebbles. Some of these stones they brought Kokorin. Kokorin says that all stones look like crystal rock. Mr. Kokorin is illiterate."

A couple of remarks. The first one is about Lakura. On modern maps Lakura is about 70 km to SW from the epicenter, and according to [Vasil'ev, et al, 1981] in 1959 it was suggested that the word "Lakura/Lakursky" is not a proper name, but that Evenks denote any forested hill as "Lakura". The second one is that there are some deposits of Iceland spar in the region, and in one of them (about 29 km to SE from the epicenter) even a small mine/placer was working some years ago.

At those times Kulik was concentrated to find meteorites near the epicenter of the forestfall, so he rejected ideas to search remote areas. Here is, for example, regarding Evenk Ivan Ilyich Onkoul who was sent to L.A. Kulik 5/VI-1930 by the Suglan (congress) of the Evenks [Vasil'ev, et al, 1981] (translated by A.O.):

> "He allegedly saw a "dry river". L.A. Kulik drew up an act stating that I.I. Onkoul refused to indicate a "dry river", saying that there is no such river, that any river at low water can be dry."

Some interesting data was obtained by I.S. Astapovich (Astapowitsch). In the course of the astronomical and geophysical expeditions of 1930 - 1932, on the rivers Angara and Lena, he gathered data regarding the event. A. Ovchinnikov (Geophysical Institute, Irkutsk), who visited the epicenter in August, 1932 shared some info with him [Astapowitsch, 1940]. Anyway, in 1951 Astapovich published a relatively large review of the event, and wrote in there the following info (the author is unaware why the Ovchinnikov's name changed from 'A.' to 'S.')[Astapovich, 1951b] (translated by A.O.):

> "The explosion caused a powerful convection of air masses, which cooled when lifted up, which led to the formation of a piniform or mushroom cloud and thunderstorms; rain poured down, which may have flooded the taiga fire that started.
> <...>

Geophysicist S. Ovchinnikov informed us that the Evenks mentioned a sulfuric smell in the area of the fall; he thinks it was ozone."

## 3. Discussion

Let's pay attention to words about rain by Astapovich. He was probably the only one from the first generation of Tunguska researchers who starts to say about the rain. Indeed after Kulik's return from the first expedition, a prominent Soviet meteorologist B. P. Multanovskii suggested that the possibility of cyclonic phenomena in the area of the alleged meteorite impact site is not excluded. However, Multanovskii pointed out, due to the lack of meteorological stations in 1908 in the area of the event, weather data is not available to the Weather Bureau.

Kulik replied to his critics in his numerous lectures and articles (in popular publications in general). For example, he wrote in a popular magazine "Vestnik Znaniya" (1927, N22, p.1356) (translated by A.O.):

"Finally, and this is the most important thing, at 7 a.m. on June 30, 1908, there was a steady anticyclone over the Yenisei basin: it was absolutely quiet from the tundra to the Sayan Mountains, and the sky was cloudless. This is evidenced by the data of meteorological stations, and the unanimous testimony of eyewitnesses, from wild Tungus to highly qualified scientists who happened to be in the area."

In reality localised thunderstorms could occur even inside anticyclone, moreover as, for example, A.V. Voznesenskii wrote in 1925 [Voznesenskii, 1925] that weather was fair in the south of the affected region and cloudy in the north.

Astapovich was also a little bit more cautious than Kulik, and while excluding "an ordinary windstorm", he recognized clouds in the north [Astapowitsch, 1940]:

"With the aim of studying the general meteorological conditions at the time of the fall of the meteorite, the author, in 1932, collected the observations of meteorological stations that made observations in 1908… <…> An examination of these data shows that throughout nearly the whole region of observation the weather was anti-cyclonic, and only in the northern part (Dudinka, Khatanga) was it cloudy; unfortunately, the lack of observations between the parallels of 60° and 70° renders it impossible to determine exactly the boundaries of cloudiness. Thus, the meteorological data well confirm the reports of eyewitnesses of clear and calm weather during the flight of the meteorite, which favored its observation. With this fact vanishes every doubt that the whirlwind personally experienced by …

and possibly by others, was occasioned by the meteorite and not [by] an ordinary windstorm."

In 1951 Astapovich continued to insist on the "fair weather" [Astapovich, 1951b] (translated by A.O.):

"Meteorological conditions for a given moment corresponding to $6^1/_2$-8 hours the mornings local time, based on data from the same 22 meteorological stations [5], were very favorable: in Central Siberia there was anticyclonic weather with temperatures from +11 ° (Ilimsk) to +24 ° C (Turukhansk), weak wind (less than 5 m/sec). These data are in good agreement with the descriptions of eyewitnesses. To the north of the crash site, about 65° parallel, there was a border of clouds, so only in Dudinka and Khatanga it was overcast. Therefore, there is no suggestion that the windfall was caused by some kind of hurricane."

In 1951 Astapovich even published a map with supposed cloud cover region [Astapovich, 1951a], which is reproduced on Fig.3. AB is the projection of the alleged bolide trajectory onto the Earth's surface, and inclined lines in the north highlight the area of continuous cloud cover on the day of the event (according to Astapovich).

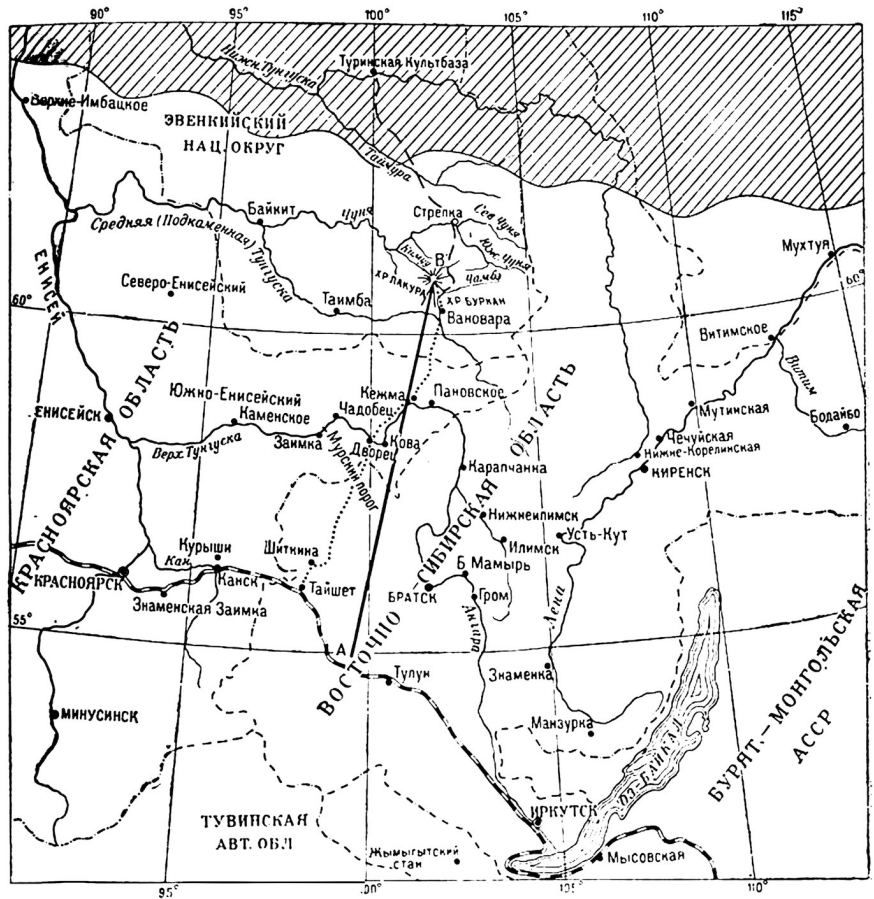

**Fig.3**

Remarkably that Astapovich marked the map as of indefinitely "on the day of the event", while even data from the meteorological station in Kezhma (which Astapovich presented in [Astapowitsch, 1940]) shows complete cloud cover at 1 p.m. Moreover at 1 p.m. cloudiness strongly increased at several more meteorological stations to the south from the epicenter.

E.L. Krinov in general repeated after Astapovich [Krinov, 1966]:

"This table shows that an anticyclone existed over almost all the region of fall; only in the extreme north was the weather overcast, hence the conditions for observing the bolide were nearly everywhere favourable, or even good. The meteorological data moreover show that the destruction of the forest and the effect of the aerial waves, which individual eyewitnesses experienced, were undoubtedly caused by the fall of the meteorite and not by an ordinary hurricane. The hurricane theory was put forward by some scientists after the results of Kulik's investigation of this region had become known in 1927 and later."

How correct are the given above weather classifications will be discussed below.

Fortunately, accounts obtained in the 1960s, as well as data from the Kezhma meteorological station (see below) allowed the author of this paper to clarify that clouds appeared much farther south in the morning of June 30.

Let's consider the data of the Kezhma meteorological station which was about 215 km to the south from the epicenter. The observer of the Kezhma meteorological station - Anfinogen Kesarevich Kokorin (1872 – 1938) was interested in the event, collected info, etc. In the late 1920s he even helped to L.A. Kulik (Kulik wrote in 1930 in a letter that Kokorin was considered by local authorities as a valuable employee, but was "lishenets" – i.e. deprived of some civil rights). Unfortunately during the Stalin's terror, Kokorin was accused of counter-revolutionary propaganda and shot. He was rehabilitated in 1958.

On the margins of the station weather log-book Kokorin wrote the following (translation by A.O.):

"June 30…new calendar at 7 o'clock in the morning, two huge fiery circles appeared in the north; after 4 minutes from the beginning of the appearance, the circles disappeared; soon after the disappearance of the fiery circles, a strong noise was heard, similar to the noise of the wind, which went from north to south; the noise lasted about 5 minutes. Then followed the sounds and crackling, similar to the shots from huge guns, from which the frames trembled. These shots lasted for 2 minutes, and after them there was a crackling sound, similar to a shot from a gun. These last lasted 2 min. Everything that happened was under a clear sky".

His account is remarkable in many aspects (absence of any super-bolide and even its trail, etc.), but here just one point will be considered – clear sky, while he marked in the weather-log cloudiness at 7 am as 4 (near the data for the cloudiness at 7 am and 1 pm there is a sign that the Sun is covered with clouds and shadows from objects are weak). Opinion of the author of this paper is the following. An upsurge of cloudiness took place right after the event or maybe even coincided with the event. Here is an account collected by a prominent Soviet geologist (and researcher of the Tunguska event) Boris Vronskiy (https://commons.wikimedia.org/wiki/Category:Boris_Vronskiy ), which he wrote down in his diary (http://prozhito.org/person/386) on July 21, 1960 (by the way, the account seems to be missed in [Vasil'ev, et al, 1981]) (translated by A.O.):

"The story of Pyotr Kallistratovich Bryukhanov, aged 62-63, an eyewitness to the fall of the Tunguska meteorite.

He and his mother were located 60 km from Vanavara in the direction of Kezhma on the zaimka or settlement. They were going to take out fish from zaezdok {a fence with traps in the water for catching fish – A.O.}. Suddenly, a luminous streak flashed through the air, which quickly disappeared behind the forest. The mother was frightened and, grabbing Petya, who was 10 years old, by the hand, ran to the hut. Suddenly there was a strong thunderclap, followed by another, a third, up to a dozen, which gradually weakened. The ground was shaking under feet. In the north, the sky was black as soot. When they ran up to the hut, the windows were broken, and the door was open."

The account is very interesting, pointing to extremely black sky in the north right after the event (or maybe even earlier).

In 1962 Vronskiy wrote down in Vanavara another interesting account in his diary on Aug.9, 1962 (at that time his duty was to search for tiny spherules from the alleged Tunguska spacebody). Unfortunately a place of the observation is not mentioned. Here is its translation by A.O.:

"Today, the old man ("grandfather") is on duty at the airfield. Unlike the women who were on duty earlier, he meticulously walks around the terminal building, inspecting the locks, and generally shows signs of violent activity. He came to me, asked who I was, what I was doing, said that he remembered the situation in which the meteorite fell – he was 7 years old at the time. The mother left early in the morning on a boat to milk cows, and the children were sitting at the open window, looking at the street. Suddenly a terrible hurricane arose, darkness thickened for a short time, the cabinet doors opened with a clang and a crash, and dishes flew out. The children began to cry. And then something flashed, and there were frequent blows, similar to cannon volleys. Many shouted "the Japanese is coming" and began to hide underground. They believed that the Antichrist was coming and the end of the world was coming. The population was so scared that a religious procession was spontaneously organized around the village. The Tunguska catastrophe sacredly keeps its secret, and our spherules are unlikely to be able to decipher it."

It is remarkable that Vronskiy (who in 1958 was a strong advocate of the fall of the meteorite) already in 1962 was skeptical about results of the search for the alleged spacebody's substance. Among the reasons could be that in 1959 he got acquainted with the geological report of discovery of nickel-iron in sufficiently significant quantities in the concentrates of the Ognia River. This place is about ~55-60 km (azimuth about 128 degrees ) from the epicenter. It is possible to add that nickel deposits were discovered also at ~ 90 km to SE from the epicenter. By the way there are also scattered deposits in the region including the area near the epicenter of Fe, Au, Cu, Zn, Ba, Sn, Pb, Na, K. The epicenter is inside the Kimchunskii feldspar-bearing area.

It is interesting to compare the "grandfather" account with a fragment of report from Kezhma published in the newspaper "Krasnoyarets" in the issue of July 13, 1908 (the Julian calendar), translated by A.O.:

"But with close observation, in the north, i.e., where the blows seemed to be heard - on the horizon, something like an ash-colored cloud was clearly noticed, which gradually decreasing, it became more transparent, and by 2-3 o'clock in the afternoon it completely disappeared."

Reports of the darkening came from several places. For example in [Epiktetova, 2003] an attempt was done to explain the darkening by a shadow of the Tunguska spacebody substance which partially dispersed (translated by A.O.):

"Calculations of the coordinates of the trajectory points have been started on the basis of accounts that indicate a darkening of the terrain that occurred both before and after the arrival of sound. This is about two dozen accounts from settlements located in a compact area bounded from the West by an approximately azimuthal beam of 220° from the epicenter of the forest fall: Sagaevskoye (220°, 1100 km), Kansk (217°, 635 km), a weak phenomenon was observed in the village of Klimino on the Angara River (217°19', 304 km). The strongest "eclipse" is described in the accounts of witnesses from Kansk and Taishet. Slightly weaker darkening was noted in villages on the Angara River above Kezhma and on the Ilim River. In the East, the area of darkening is bounded approximately by the upper course of the Lena River."

It is noteworthy that, according to Epiktetova [Epiktetova, 2012], information about darkening was obtained by chance during a survey of eyewitnesses in the 1960s - there was no such question in the questionnaire, and eyewitnesses reported darkening on their own initiative. The words about the darkening came as a surprise.

According to calculations by L. Epiktetova [Epiktetova, 2012] the dispersion (which shadowed the Sun) took place even at altitudes as high as 1300 km at least. The author of this paper does not comment such explanation of the darkening.

Darkening was also reported by Evenki from places not far from the epicenter in accounts collected in 1960s. Here is from [Vasil'ev, et al, 1981] (translated by A.O.):

"Aksenova O. was interviewed in 1960. In 1908 she lived in the upper reaches of the Mutorai River. At that time she was 20-24 years old.

"The weather was fine early in the morning, then the wind went, it got dark, as before the rain, the ground turned red and the thunder went strong. I didn't see anything in the sky. According to rumors, there was a strong fire on Khuzhma {apparently the Khushma river - A.O.}. After that, people were very ill with smallpox. We lived far away and didn't go there.""

Aksenova O. was about 54 km at azimuth ~ 244° from the epicenter. A red color is remarkable. It is noteworthy that K.P. Florensky (who visited Tunguska in 1953) mentioned about Evenki eyewitness who remembers "how the sky burned", but Florensky added that such accounts are not of interest now [Ol'khovatov, 2020a]. Fortunately some other Tunguska researchers (especially starting since 1960s) did not ignore such accounts, but often it was already too late…

There are several more Evenki accounts about reddening. Here is a couple accounts collected probably in 1964 and taken from [Vasil'ev, et al, 1981] (translated by A.O.):

"Elkina Anna Yakovlevna, 75 years old, an Evenk living in Vanavara. Place of observation: the Northern Chunya {river - A.O.}, 30-50 km. from Strelka-Chunya on the Ananyakit creek.

"In the morning, early - early at 5 o'clock, it thundered {from – A.O.} a little higher than the sun. High – high above. The whole sky was red, and not only the sky, everything around was red - the earth and the sky. Then there was a strong thunder. The sound is like bells, like beating on iron. There was thunder for half an hour. When the thunder started, the red disappeared (it disappeared all at once, at the same time). The earth was shaking violently, the chum was staggering. There was no wind, only thunder, and the ground was shaking. There was no smoke. The morning is clear - clear. The sun has already risen a little."

Was there a ban on going to that place? No, she says, they went at the same time, the same autumn."

Unfortunately the author of this paper failed to find the creek on a map. Anyway it is possible to say that Elkina was somewhere ~ 120-140 km at azimuth ~ 30 – 40 degrees from the epicenter. Again remarkable aspect - "everything around was red". And the redness disappeared with after the thunder started.

Let's look at one more account taken probably in 1964. The eyewitness was about 76-80 km at azimuth ~316 degrees from the epicenter. As it can be seen from the account, clouds were already presented in the area on that morning. Here is the account [Vasil'ev, et al, 1981] (translated by A.O.):

"Dmitrieva Maria Vasilyevna, Evenk, aged 96 years, living in Strelka-Chunya, clearly remembers the events associated with the disaster.

According to her, at the time of the disaster, they were camped close to the mouth of the Kimchu, but north of the river itself.

There were 6-7 large birch bark chums in the camp, in which there were about 50 people.

The weather this morning was not rainy, but the morning was gray, the sky was covered with high clouds, there was no wind. It was early in the morning, and many were still asleep, but some had already got up and were outside of the chums.

There was an explosion. During the explosion it was like an earthquake.  Some chums were blown away by the wind and had to be held with their hands, some chums resisted. At the same time, several strong sounds were heard, different from thunder, more prolonged and sonorous {ringing - A.O.} (bynn - bynn - bynn).

On the question: "Which of the phenomena was noted first?", repeats that everything happened at the same time. The air wave slightly stunned the people who were outside the chum, and tore off the birch bark from several chums, which the people outside were trying to hold with hands. However, some of those sleeping in the chums did not wake up and did not hear anything. After the thunder and the whirlwind, there was silence. The deer quieted down by the smoke-holes and stood motionless.

The sounds and the wind came from the southeast. After that, the redness appeared after the explosion, the sky was red for a long time, as the dawn is red, and this redness gradually went to the west, where it was still visible for some time as a glow.

The explosions were far away. There were, by the sound, several of them. Trees were not felled. The explosions were very prolonged and did not sound like thunder. There was no fire {forestfire – A.O.} nearby, and the explosion was not on the ground, but high up. No column of fire or smoke was seen in the direction of the explosion site, but everything was happening high in the sky, from where the sounds were coming. (The old woman pointed under the lintel several times, at 60 degrees).

After additional questions, I. {probably mistype – must be M. – A.O.} V. Dmitrieva repeated that everything happened in the side of the lunchtime sun, at altitude. The redness in the sky appeared after they heard the sounds."

There are several interesting aspects of the account besides cloudiness: a) the redness was despite clouds which is an argument that it was presented at low altitude; b) the sequence the thunder/explosion – the redness is reverse compared with the Elkina's account - can't it be due to difference in distances and positions in general? c) The redness gradually went to the west. It can't be ruled out that the "redness gradually went to the west" is due to the rising of the sun, when the eastern part of the sky becomes lighter and the redness is less noticeable against this background. Anyway here is a story from the west collected in 1969 [Vasil'ev, et al, 1981] (translated by A.O.):

"Chicherin Anton Nikolaevich, born in 1907, Russian, higher education.
Transmits the stories of grandfather Innokenty and grandfather Lazar, direct eyewitnesses of the Tunguska disaster.
In 1908, grandfather Innokenty lived in the village of Podkamennaya Tunguska or on the Chyornaya rechka (a left tributary in the lower reaches of Podkamennaya Tunguska).
There was no explosion, but there was a rumble with a concussion. There was a red glow in the east for a long time. Grandfather Innokenty often compared the fall of the "Tunguska miracle" with a terrible thunderstorm of extraordinary strength, which took place on a wide front in July 1932. The thunderstorm glowed in red. Interviewed in the village of Podkamennaya Tunguska."

The village of Podkamennaya Tunguska is 637 km at azimuth 282 degrees from the epicenter, and the Chyornaya rechka is 572 km from the epicenter at almost the same azimuth.
Nevertheless, the opinion about "fair weather everywhere" was so well-established among most Tunguska-researchers that when the first Evenki accounts of rain were received in the late 1950s, it caused confusion. Here is an example of such an account. Evenk woman Nastya Dzhenkul ( the widow of Ivan Maksimovich Dzhenkul ) informed (among other things) in  November 1959 [Vasil'ev et al., 1981] (translated by by A.O.) that:
"Her father and grandfather lived at that time (in 1908) on the Khushma River. The weather was good, suddenly it began to rain, a strong wind rose, and dragged away the birch bark tent. A large stone fell, as big as a tent, jumped two or three times, and then drowned in a swamp. The stone was shiny, black, fell with a terrible sound - u - u -u -u."

Another account (taken in 1965) from [Vasil'ev et al., 1981] translated by A.O.:

"Masmoro Trofim, born in 1888. Born and raised in the vicinity of the village of Taimba (nomads). When he was young (it was not possible to tie him down more precisely in time), he roamed with deer in the upper reaches of the Podporozhnaya River, about 2 km below the Miryuginsky threshold, 40 km from the mouth on the left tributary.
In the morning there were clouds with rain and it was dark as night. There was thunder, sparks flew. There was a very strong wind, felled trees and killed 2 deer. The deer ran away from the smoke hole. The thunder was with the stronger last three strikes. The last one was especially strong ("boom-boom-boom!"). The ground was shaking. Clouds and rain were up to the last blow, then it became clear again. The direction of the clouds is the same as the body. The body was moving approximately at an angle of 30° to the magnetic meridian (the direction was indicated several times unambiguously), almost at the zenith. Flew like a rocket. The tail is weak. And then it fell (this thought was repeated many times. It seems that it was not the flight, but the fall that was the most vivid and strong impression). The old men then went to look. Nothing was found. They said "fire fell into the Tundra." In the upper reaches of the Podporozhnaya River, {he - A.O.} saw a fall of forest on the ridge, which had not been there before."

The place of observation is about 126 km and azimuth 256 degrees from the epicenter. Unfortunately the part of the account with "the body" is not very clear. Anyway there are several interesting points (besides clouds and rain, which ended after the last blow). "The body" was seen near the zenith, and it looks like under clouds. There were seismic phenomena, and probably a forestfall at distance more than 100 km from the epicenter. If even to admit that the event described by the witness is not the 1908 Tunguska event, then it means that another peculiar Tunguska-like event took place not far from the epicenter at those times.

In 2018 an extended electronic version of [Vasil'ev et al., 1981] was published in internet (see: tunguska.tsc.ru/ru/science/1/eyewitness/ ). Here is an interesting account written down by (apparently) N.V. Vasil'ev in 1970 and translated by A.O. (although, strictly speaking, there is no absolute certainty that the description refers to the Tunguska event, nevertheless it is very remarkable, and it resembles some other accounts):

"An account on a diary sheet (apparently, N.V. Vasil'ev entry).
18.08.70. Zakhryapin ... Innokentievich, born about 1898. In 1908 he lived in Vanavara, at the trading post. He was then 8-9 years old. He remembers Semenov well ("Borisyata") and Kosolapov. (They were later dispossessed {dekulakized - see https://en.wikipedia.org/wiki/Dekulakization - A.O.} and exiled from Vanavara, he does not know their further fate).
It was early in the morning. A black cloud swooped down. It took about 10-15 minutes. There were very strong thunderclaps. It started raining heavily. In fright, he went underground. Adults who were outside the room were thrown in different directions.
Later, in the 30s {1930s - A.O.}, he was well acquainted with Kulik, accompanied him to Baykit when he {i.e. Kulik –A.O.} was rafting to the mouth of the Podkamennaya Tunguska. He confirmed that the oldest buildings in Vanavara stood approximately in the area of the present building of the district bank, and not on the cliff itself. Regarding the old Ilympeiskaya road, he said that this is the so-called Yuktinskaya - a road well used by current hunters as well.

The old man makes a very pleasant impression, speaks quite logically, has a clear, distinct memory."

On Fig. 4 there is a photo of N. V. Vasil'ev near the Cheko lake in 1979 (from the left to the right -  E.V. Lodkina, N.V. Vasil'ev, B. M. Trubetskoi - thanks to E.V. Lodkina for her permission to use the photo).

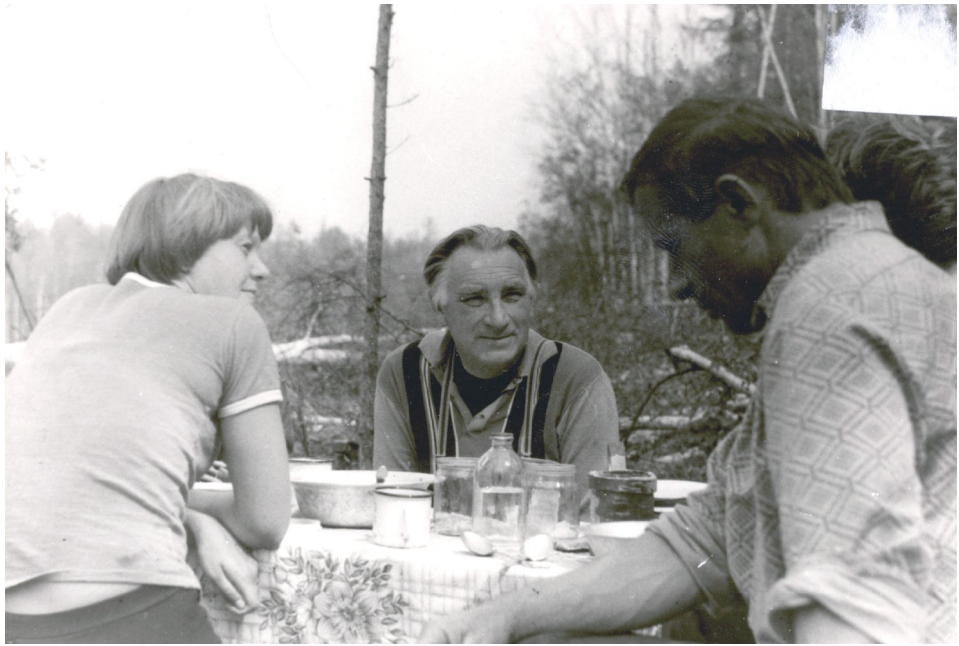

**Fig.4**

The first time N. Vasil'ev wrote down about rain from a witness was in 1959. In 1959, before the first KSE expedition, folklore writer Ivan Ivanovich Suvorov advised KSE to meet with an eyewitness Ivan Alekseevich Ayulchin. Here is from the Vasil'ev's diary ( https://tunguska.tsc.ru/ru/science/mat/oche/92-120/68/ , translated by A.O.):

"Conversation with Ayulchin: "He is 102 years old, he is now in a district hospital. The old man is quite cheerful, remembers the previous events well. Regarding the fall of the meteorite in 1908, he tells the following: at that time he was on the Northern Chunya. Early in the morning, a terrible roar was heard, resembling the blows of a huge tambourine. The wind rose, the bark was torn from the chums. The Evenks, frightened, wrapped themselves in a blanket and prayed to God. At that time there were Evenks from the genus Kainachenok on Khushma, who were looking for their warehouses there; some people died during the explosion, some got out alive, these people for a long time lived after that. There was rain after the explosion (?). Fish disappeared in Khushma for a long time, hunting that year was also very unsuccessful."

It is remarkable that apparently astonished Vasil'ev placed sign "?" after rain.

Ayulchin was no closer that 106 km from the epicenter, sooner about 120-130 km to ~NE.

The meeting with Ayulchin was recorded by a professional journalist Aleksandr Stepanovich Erokhovets. He took part in the KSE expedition. In 1960 he published in the "Sibirskie ogni" magazine a documentary story about the expedition. He is a fragment from it concerning the meeting with Ayulchin (translated by A.O.):

"The next day we visited the Evenk Ivan Alekseevich Ayulchin, who is in the hospital, who is nicknamed the "white old man". Ivan Ivanovich Suvorov, the director of the Museum in Tura, advised us to meet him back in Krasnoyarsk.

<...>

The old man — thin, bent, with a white head — was taken out of the hospital by the arms. Ayulchin slowly shifted his skinny legs, shod in soft boots made of deerskin — bokari.  He has an intelligent face with wrinkled hollow cheeks. The convex forehead is also pitted with large wrinkles. White hairs on the head were sparse and light as fluff.

<...>

Cautious inquiries began. But they didn't give much. The old man remembered that after the explosion, the yurt suddenly fell. The sky was glowing. Near Khushmo, the thunder was like a tambourine which beat: butym-butym-butym-bom!  Then it started raining, it rained all day. All the Evenks prayed to God — what an end of the world!..

— Amana, did you go to the place of the explosion?

— I did not go. Hunters were afraid to walk. Deer were burning, people were burning on Khushmo. The forest-fire is coming — God forbid. There was a rich labaz there, Kainochenok went there, one burned down, two ran away."

From the text it is seen that Erokhovets was a bit disappointed: "they didn't give much". Indeed two other witnesses whom they met said about the event as a remarkable one, but without such things like a super-bright phenomena in the sky (or just super-bolide trail at least), etc. Anyway, here are fragments from the Erokhovets

story with the accounts (taken in Kezhma) and translated by A.O.:

"On the shore we met an old fisherman who remembered the time when the Tunguska meteorite flew by. It excited us. However, we could not find out anything interesting from him. The old man said that he had seen the meteorite fly by. He was living near here at that time. At first, smoke appeared, then a terrible wind swept through, even the trees were bent to the ground. People got scared and ran to the village. His neighbor had two windows blown out by the wind."

Please pay attention that Kezhma was a rather large village (1248 residents on Jan.1, 1911) stretching along the river in appr. E-W direction. So a low-flying object seen at one end could be missed at another end.

The second account is described with more details. The talk is between KSE and probably Osip Efimovich Kosolapov (born 1875) who seems to be rather cheerful (translated by A.O.):

"— Do you remember how a meteorite flew by in the year 908? Did you live here then?
— And {I – A.O.} lived in Mozgovaya. This is such a village, about 12 versts from Kezhma, upstream... How, I remember, I remember. I was still strong then, y-yes... we were on arable land at that time... Together with Kirik Kosolap, such an old man, my namesake. This is a nickname, but he was really called Kiryan Markovich Kosolapov. And so. We sat down, then, to have tea, suddenly I hear: uh-uh, hook, hook, hook! Like shooting. I thought that the Kezhma's hunters had arrived, they were shooting at bears in the taiga with big guns. And Kirik says: "Thunderstorm!" And the weather was quiet... Only the sky is red. But on Katanga there was a strong wind, they said that the air was just coming, trees were felling.
— And you weren't scared?
— And why should you be scared? We drank tea and went to plow again. The family then, however, said that the cups on the tables were shaking and the glass in the huts, which was weaker, was pulled out by the air... "

Mozgovaya is about 214 km at azimuth 190 degrees from the epicener. So there was red sky at such distance too. But no superbolide or just its trail at least. Indeed, for example, I.S. Astapovich had to admit that [Astapovich, 1951b] (translated by A.O.):

"We had several tens descriptions of eyewitness accounts at our disposal, such as "suddenly a huge fiery mass surrounded by a glowing atmosphere appeared" [54], "some kind of fiery celestial body cut through the sky from south to north" [56], etc. A careful examination of these data showed that at the beginning of its flight the bolide was "as bright as the Sun", and by the end of the journey "many times weaker than it", "it was possible to look at it", although the bolide approached the surface of the earth. It can be thought that the bolide has reduced its brightness by hundreds of times by the end, from -28 to -21 stellar magnitude (the brightness of the Sun is -26.7 st. mag.)."

But as it can be seen from accounts (see also [Ol'khovatov, 2022; 2023]), many eyewitnesses (who were in good positions for observations) did not see any bolide at all (as well as its alleged trail) or saw some object flying in various directions. The author of this paper can say that there is just a minor percent of eyewitnesses who said that the bolide was very bright.

Now let's consider the Astapovich's explanation of the rain which was presented early in this paper:

"The explosion caused a powerful convection of air masses, which cooled when lifted up, which led to the formation of a piniform or mushroom cloud and thunderstorms;…"

If we do not go into the details of the Astapovich's explanation, then the following should be noted: a) for the implementation of such a mechanism, it is necessary to have a sufficient amount of moisture in the air and other conditions; b) the explosive effect serves only as a trigger that triggers the corresponding atmospheric processes. Thus, the atmosphere for a thunderstorm should be "ready", and the explosion itself is just a trigger mechanism.

By the way, the "dark clouds" and just darkening of the sky were observed on a very large area, where the mechanism proposed by Astapovich would have been ineffective, or even absent. Moreover research revealed that the associated forest-fire intensity was rather moderate or even low in general (with exception in the area near the epicenter) while initially starting over rather large area. Of course, in some places the forest-fire could be stronger and/or last longer than in others, but in general it was rather moderate. Astapovich already admitted in 1951 (see early in this paper) that "rain poured down, which may have flooded the taiga fire that started". Later researchers also gradually came to conclusion that the forest was wet. Here is from [Doroshin, 2005] (translated by A.O.):

"The extinction of the forest-fire at the place of its occurrence indicates poor conditions for the spread of the forest-fire. Indirectly, this is confirmed by the features of the "bird's claw" type of lesion (perhaps the catastrophe was preceded by either light rain or heavy dew, see above).

Practically throughout the entire territory of the fire, with the exception of its central part, only a weak forest-fire can be traced."

As studies have shown [Gorbatenko, 2003], in June 1908 in Kezhma there was a record amount of precipitation compared to all previous years of observations. But on the other hand, the last rain (in June 1908) in Kezhma was registered on June 25 and it was very light rain. A very light rain also was registered in Kezhma on July 1 as well as a distant thunderstorm.

Authors of another forest-fire research [Yashkov, Krasavchikov, 2008] wrote (translated by A.O.):

"The already available results and field observations allow suggesting the following:
a) the best analogy for the mechanism of fire can be the spraying of a burning substance on a bonfire that has not yet been lit, made of wet wood: a strong fire at the place of fire, subsequent attenuation and the absence of prolonged burning; <…>"

Remarkably but something similar was written as early as in 1929 [Smirnov, 1929] (translated by A.O.):

"The trees bear traces of a severe burn, but the real burning is nowhere to be seen, and this involuntarily stops attention. Why was there no forest-fire in the taiga, since there was a fire here? True, it is difficult to light raw trees standing on the root, but there is enough dry deadwood in the taiga, which sometimes lights up from a cigarette butt. The explanation of this curious phenomenon, in my opinion, can it can be twofold: either the burn was so lightning fast that it could not cause a real fire, or the fire that started from the burn soon stopped. Residents of the Angara region, who remember the fall of the meteorite, say that after the fireball fell into the taiga, a strong, unprecedented thunderstorm with a downpour soon broke out."

The author of these lines was Aleksei Smirnov - an experienced traveler, a hunter, a local historian and a writer. In 1928, he participated in a rescue expedition aimed at helping/saving L.A. Kulik.

Unprecedented level of the thunderstorm hints that the event was indeed very peculiar. By the way possibly it can explain info collected by S.V. Obruchev in 1924 (see early in this paper) that the rumble was heard in the morning, but according to other accounts in about 10 hours.

That's a pity that among the first generation of Tunguska researchers only I.S. Astapovich mentioned the thunderstorm, and only in 1951…

In 2008 a meteorologist from Tomsk (who took part in the KSE expeditions)

presented results of a new meteorological analysis, which says (among other things) that the meteorological situation was rather complicated. Here it is [Gorbatenko, 2008] (TKT is the alleged Tunguska space body, translated by A.O.):

"Judging by the changes in air temperature and pressure at stations located around the place where the TKT fell and during the fall, it can be assumed that over the territory of the south-west of the Eastern (and partly in the southeast of Western) Siberia there was an extensive cyclone. Its warm front sequentially passed through the settlements: Kansk, Yeniseisk, Kezhma. The center of the cyclone was located east of Krasnoyarsk, in the Kansk region, with a pressure in the center of about 735 mm Hg (980 mb).

In addition, analyzing the air temperature values (Fig. 2) at stations close to the area of interest to us, it is safe to state that in the period from 06/30/1908 to 07/01/1908 above the territory of interest to us passed a warm front and for 0.5-1 days the territory was in a warm sector of the cyclone, and therefore at this time there could be stratus clouds and a general deterioration visibility. A little later it was replaced by a cold front with cumulus clouds and thunderstorms."

So the idea of Boris Multanovskii about "cyclonic phenomena" seems to be close to reality. Anyway, for example, a weather log of the Kirensk meteostation (situated about 491 km to the SE from the epicenter) marked in the afternoon of June 30 a strong wind. For comparison, the previously strong wind was noted by this meteorological station in 1908 only on March 31, and next time - December 9th. So June 30, 1908 was a peculiar day in the region indeed…

Now let's consider the Evenki accounts collected by Suslov in 1926.

From the Akulina's account, the following sequence of effects emerges approximately:

1) mechanical impact on the chum (probably, seismic and/or wind). Judging by the description, the seismic one, with a high probability, was present. Perhaps then the wind effect also joined it (it can't be ruled out from the account that there could be some lights/flashes while Evenks were sleeping).

2) Bright light and strong wind. From the account it looks like duration of the light was no shorter than 1-2 seconds at least.

3) Sounds reminiscent of shooting.

4) The impact of the whirlwind.

5) Burning of dry things (wood, etc.) in general (but this may start earlier).

Regarding the moderate forest-fire reported by Akulina, it is reasonable to say that authors of [Yashkov, Krasavchikov, 2008] (who researched the fire damage in Tunguska for many years) discovered traces of the associated fires at distances more that 20 km from the epicenter, and they came to conclusion that eyewitness accounts of forest-fires in areas far enough from the epicenter are not a fantasy.

Interestingly that a rain was not reported in the Akulina account. But near the epicenter a large amount of heat was deposited and the rain could be absent in there indeed during the event. But there is a minor hint to a past rain in the Akulina's words (when she moved away from the epicenter) that the fire diminished, and it became cold (but of course, there could be another explanation for the words, as torn clothes, etc.).

One more interesting aspect - Ivan was thrown away for about 40 meters by a whirlwind. Indeed, it couldn't have been a blast wave, as in the Ivan's case the manifestation was localized (on Ivan).

Now let's consider the account by brothers Chuchancha and Chekaren. Approximately the following sequence of events emerges from the brothers' story:

1) A mechanical impact on the chum. Judging by the description, there was a wind impact and repeated seismic impact.

2) The first thunderclap.

3) Constant seismic impact and the wind that brought down the chum.

4) After about 6 seconds after the first thunderclap, a lightning-like brightest flash and then at once the second thunderclap.

5) The next lightning-like flash and immediately the third thunderclap, followed by wind.

6) The next flash and the fourth (less severe) blow. Judging by the account, the duration of the flash was at least a several seconds.

7) The fifth blow somewhere far away to the North-West.

Unfortunately, at this point the account ends.

Judging by the account, duration of the flashes was much longer than discharges of an ordinary lightning.

It is interesting that flashes and thunderclaps were practical simultaneously, and fast wind actions are remarkable too. It resemblances the event described in [Worth, 1972]:

"I was standing on top of the Puy Mary, 1,787 metres high, the highest of the volcanic cones in the range of the Cantal mountains. To the north, about 1-2 miles away, I saw lightning coming out from a black bank of clouds and heard the thunder about 5-6 s later. However, I felt a strong blast of hot air reaching me before I could hear the thunderclap. This happened several times in quick succession, though I did not time the length of the intervals between each lightning."

Now let's consider the account by Ulkigo. The beginning of the description of the event (suddenly, early in the morning, the dogs howled, the children cried) is very close to the description of the beginning of the event from an even more remote point - the settlement of Sulomay [Ol'khovatov, 2020a]. The Ulkigo's account demonstrates that seismic phenomena were first to start. Some of them coincided with the thunderous sounds. A sky-flash and the thunderous sounds were almost simultaneous.

Besides the presented Evenki accounts, Innokentii Suslov in a letter to a journalist wrote in 1961 about his participation in the Evenk's Suglan in 1926 ( http://tunguska.tsc.ru/ru/science/mat/oche/31-60/d-041/ , translated by A.O.):

"The article "Agdy" contains detailed stories of the Tungus (Evenks) who flew above the forest, saw a burning forest, smoke under them, survived the horror of the roar, blows and were on the verge of losing their minds. They spoke in detail about the details of the disaster.
<…>
The experience was a success. It challenged the Evenks to frankness at such an authoritative meeting. As a result, all the stories of individual victims (who flew) were confirmed here... "

Unfortunately to the author's knowledge, the article "Agdy" was never published, and the "flights" are absent in (known to the author) Tunguska accounts presented by Suslov. The reason for this is unclear, but maybe poor health of I. Suslov's wife (who typed his notes) was involved. Also possibly it was because to some reviewers these Evenks accounts seemed "too incredible", and Suslov (having problems with health) later decided not to insist and excluded them. Similar reaction happens regarding some accounts of Tunguska – see early in this paper. Moreover authors of [Vasil'ev,et al., 1981] considered the accounts to be very reliable also because they were collected during Suglan, when telling a lie was considered to be a serious misconduct.

Please pay attention that the reported Evenk's flying seems to occur some time after starting fires, as the forest-fires seem to be rather developed at these moments. It is in agreement with Evenki "on the ground" accounts some of which point that a windstorm continued for a relatively long time.

It is remarkable that statistical analysis of about 700 accounts (collected in various years) conducted in [Demin et al., 1984] revealed that (translated by A.O.):

"The Tunguska phenomenon, according to eyewitnesses, was accompanied by various meteorological phenomena (Table 7). The most frequently observed were "strong wind", as well as "haze, fog, fog". The complexity of atmospheric processes is evidenced by sharp changes in air temperature recorded in a number of accounts. Thunderstorms, individual lightning discharges, local development of windstorms, hurricanes and whirlwinds are noted."

The above facts allow offering explanations for several unusual manifestations of Tunguska. Some of the Tunguska eyewitnesses reported a moving pillar (sometimes a "sheaf"). Here is a couple of examples. Here is the first one. Boris

Vronskiy wrote in his diary on July 18, 1970 (translated by A.O.):

"Then there was a meeting with Ivan Alekseevich Savrasov, who worked for Kulik as a driller. According to the stories, a black smoke "pillar" flew from SE to NW with a thundering (gun) fire, which then began to sparkle."

Savrasov probably drilled bogs during the last Kulik's expedition to Tunguska in 1939. The second example is from [Vasil'ev et al., 1981] translated by A.O.:

"Bryukhanova A.G. Lived in Kezhma in 1908. She was 84 years old in the year of the survey.
Interviewed by G.P. Kolobkova in 1960 in Vanavara.
"Where the sun was rising, a red pillar (of fire to the ground) passed by. Everyone just got up, the stoves were lit. I went to the basement and there I heard thunder, I thought that the neighbors were moving the hut.
The ground was shaking. I got out of the cellar, and a grandfather was lying senseless on the ground, he was thrown by the air.
The windows were intact. Then they told me that a big bright pillar of fire had gone up, there was no smoke. When the pillar fell to the ground, the ground was shaking. The windows rattled. <…>"

It is known that sometimes a funnel of a tornado glows. There are some photos – see [Vaughan, and Vonnegut, 1976; Corliss, 1982] and reference in there. Also revolving columns of fire and smoke that sweep destructively along the surface of the ground are known [Corliss, 1982, 1983]. While their physical mechanism could be debated, its discussion is beyond the scope of this paper. In this paper a phenomenological approach is used. In other words, the very fact of the existence of such phenomena is important. This could explain at least some columns/pillars reported in Tunguska accounts, including the fiery pillars seen from Tetera.
The tornadoes sometimes are accompanied by ball-lightnings. The unprecedented level of the thunderstorm in Tunguska hints too that the event could be accompanied by ball-lightnings. Here is an example how one manifestation of a ball-lightning action can explain some peculiarities of Tunguska.
On November 30, 1984, an unusual phenomenon was observed in the village of Goltsovka (now called the village of Galtsovka, 51.07 N, 82.34 E). At about 19.30, over the so-called Ryazanskaya Ridge, a luminous fiery object was seen which swiftly approached the village. The central part of this object shone with a bluish-violet light, the edges were yellowish-red and sparkled. The path of the object was not a direct line. According to eyewitnesses, the object changed its shape from ball to ellipse. The phenomenon occurred during a sharp change in meteorological parameters (a wind and a temperature). There was some moderate damage in the village resembling as from a moderate tornado. However in one place concrete pillars (60 x 50 cm) were cut off at the base and broken again. So the acting force was in this case very large, but short-ranged (the broken pillars were not thrown far away). By the way, tornadoes sometimes demonstrate similar effect [Corliss, 1983]. This allows explanation on the phenomenological level the following phenomenon related with disrupted tracheids of trees near the epicenter of Tunguska. Here is from [Vaganov, et al., 2004]:

"Given the observations of disrupted tracheids, we can make some preliminary estimates of the forces acting on these trees at the time of the Tunguska event. <…> Experiments and theoretical calculations of the tensile strength and elasticity of tracheids under tension (Mark, 1976) showed that the normal failure stress … of cells that have only middle lamellae and primary walls reaches 58.8 MPa in air dry conditions (Table 10-2 on p. 249) and is significantly less (i.e., probably 9.8-19.6 MPa, p. 33) for wet conditions. This last value, probably most relevant to the condition of tracheids on June 30, 1908, is between 20 and 30 times greater than that needed to fell the trees (Zolotov, 1961, 1967; Zotkin and Tsikulin, 1966). It is clearly difficult to explain the disruption of tracheids in surviving, standing trees near the epicenter from the point of view of an isotropic blast wave front. <…>

Tracheid transformation could have been caused by other factors. For example, live trees were found close to the explosion epicenter that had vertical or twisted stem cracks of varying depths differently oriented relative to the epicenter (Zenkin et al., 1963). These may be scars produced by lightning strikes associated with the Tunguska event."

Interestingly that scientists who investigate tree-damage appeal to lightnings, i.e. to the thunderstorm.

The (associated with Tunguska) weather phenomena could explain an interesting account by Evenk Ivan lvanovich Aksenov. Here is from [Vasil'ev, et al., 1981] translated by A.O.:

"<…> Aksenov was 24 years old. Early in the morning, Aksenov went hunting for a moose, shot a moose on the Chamba {river –A.O.} somewhere above the mouth of the Makikta {river-A.O.} and began to skin the carcass. When he was working (bending over the carcass) "suddenly everything turned red." He was scared, threw up his head - "and at that moment it hit," and he lost consciousness for a while. "When I woke up, I saw it falling all around, burning. Don't believe, Viktor Grigoryevich [V.G. Konenkin], that God was flying there, the devil was flying there. I raised my head and saw the devil flying. The devil himself was like a chock, light in color, two eyes in front, fire behind. I got scared, I closed up with his clothes on, and began to pray. (I did not pray to a pagan god, I prayed to Jesus Christ and the Virgin Mary). I prayed and woke up, nothing happened anymore. <…>".

Here is from [Vasil'ev, 1994] (TSB is Tunguska Space Body):

"Another report was conveyed to the TSB investigators by V.G. Konenkin, a local inhabitant and school teacher of physics in Vanavara, who had questioned old residents of settlements of the upper reaches of the Nizhnyaya Tunguska and the Tunguska-Chunya Region of Evenkiya for several years. Among those questioned was an Ivan Ivanovich Aksenov, an elderly Evenk man, who was said to have been a shaman hiding for many years in taiga from the authorities. The entire record of I.I.Aksenov's account is presented in Ref. [4]. At the moment of the catastrophe the eye-witness was on the Chamba river, hunting near the mouth of a tributary of the Chamba, that is some 40 km to the south from the catastrophe epicenter. A particular feature of Aksenov's account (agreeing otherwise with the early evidence of Vanavara residents that Kulik had heard as far back as the 20s), is the assertion of the eye-witness that after the explosion he had seen an object flying down the Chamba, i.e. generally north to south. He called the object a "devil".

"As I came to myself, he told Konenkin, I saw it was all falling around me, burning. You don't think, Viktor Grigoryevich [V.G.Konenkin], that was god flying, it was really devil flying. I lift up my head - and see - devil's flying. The devil itself was like a billet, light color, two eyes in front, fire behind. I was frightened, covered myself with some duds, prayed (not to the heathen god, I prayed to Jesus Christ and Virgin Mary). After some time of prayer I recovered: everything was clear. I went back to the mouth of the Yakukta where the nomad camp was. It was in the afternoon that I came there..."

Afterwards I.I. Aksenov repeated his account in the presence of V.G. Konenkin and V.M. Kuvshinnikov, an active participant of the Tunguska expeditions. In this case, however, he said that he had seen the "flying devil" not during hunting but in the afternoon, when already in the camp near the mouth of the Yakukta, also a tributary of the Chamba. The devil was going flying southward along the Chamba. It was going faster than airplanes now do. While flying, the "devil" was saying "troo-troo" (which were not at all loud).

Later, during repeated questioning in Vanavara, he did not insisted on having seen the "devil", repeating nevertheless the other evidence. When evaluating Aksenov's account it should be borne in mind that the eyewitness regarded the expedition people with distrust, considering them representatives of the "authorities", and thus the contact with him was not at all easy. On the contrary, his relations with V.G. Konenkin were fairly confidential, the latter being a local resident and a half-caste. Therefore, in our opinion, the first version appears to be more authentic, because Aksenov does not seem to have had reasons to lie to Konenkin.

What is the true meaning of this queer story and how trustworthy is it, it is now hard to say. Without overrating the significance of individual eye-witnesses' evidence, note nevertheless that at least two reports provided by those (very few) eyewitnesses who were close to the epicenter of the Tunguska explosion are really peculiar."

In the opinion of the author of this paper, Aksenov was about 30 km (or even a bit less) from the epicenter at azimuth about 165 degrees (or even a bit less). On Fig.2 his position is marked as the number "4" in red by the author of this paper. Please note that the position is shown approximately.

In 1911 when the Shishkov geodetic expedition (https://en.wikipedia.org/wiki/Vyacheslav_Shishkov ) found itself in a desperate situation in the mid of the wild taiga due to the coming winter conditions, Aksenov was one of those guides (pathfinders) who led the expedition to a safe place.

The author of this paper discusses the Aksenov story with V.K. Zhuravlev, who was involved in the communications with Aksenov in Vanavara. He admitted that Aksenov (an old sick and almost blind man with the shaman's past) was probably pressed too hard with questions. On Fig.5 there is a photo (by Kuvshinnikov V.M. from tunguska.tsc.ru) taken in Vanovara in 1969. Aksenov is to the left and Zhuravlev is near-by. The author of this paper knew Zhuravlev (unfortunately he deceased in 2020) as a very honest, kind and gentle person, so probably he was not the most disturbing person for Aksenov.

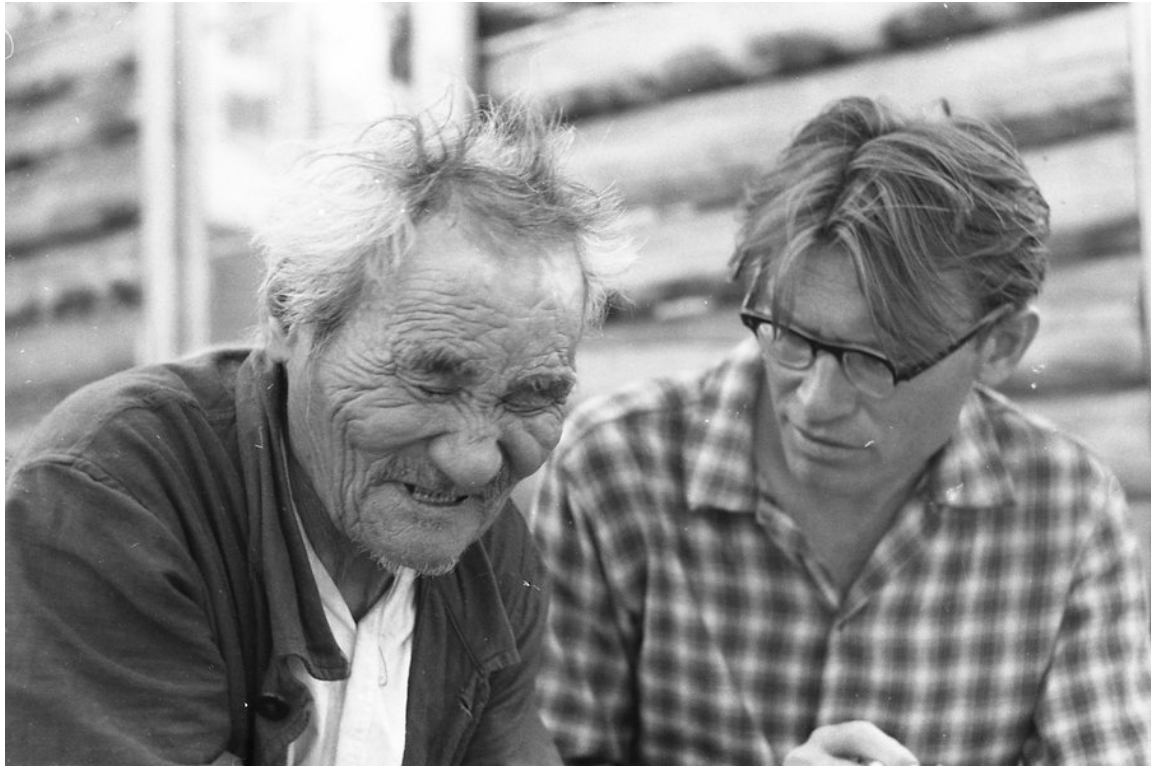

**Fig.5**

Vasil'ev was able to put Aksenov into the Vanavara's hospital where he spent last years of his life.

Could the object seen by Aksenov be one of the "Agdy birds"? Or could it be object like the black smoke "pillar" reported by Savrasov? Or maybe they were similar? Or may be similar to the event that occurred exactly one month after Tunguska? The newspaper "Sel'skii Vestnik", number 163 for 1908 (translated by A.O.):

"From Kozlovskii u. {yezd~ county – A.O.}, Tambovskaya gub. {guberniya ~ province – A.O.}, they write:

"A peasant of the Zavoronezhskaya slob. {sloboda ~ settlement – A.O.} Nikifor Matveev Sheludyakov says that on July 17, around 6 p.m., returning home with his family from the field, he witnessed the following phenomenon. On a completely clear and transparent sky which happens after a rain, two clouds appeared, one blue, the other yellow. Something small and black, like a piece of soot or ash, separated from them and, swaying from side to side, flew down. Having reached the cross on the village bell tower, this piece suddenly burst with a terrible thunder, and as smithereens splashed in all directions, one of which hit Sheludyakov slightly in the forehead, and a blue light flashed in front of his eyes, like a lit match.

Sheludyakov's horse, frightened by a strong thunderclap, ran at full speed until it hit the wall of the house. When the family of Sh-va {Sheludyakov - A.O} came to their senses somewhat, it turned out that they were all deaf, which lasted for quite a long time. There was a small red scar on the forehead of Sh-va {Sheludyakov - A.O.}. This lightning strike had no special consequences for the village.""

The latter event seems to be a good candidate for the "Agdy bird". What was it? A rare black ball-lightning? Who knows…

By the way, why did Evenki call the "birds" as "iron"? Maybe because they produced metallic-like sounds? Or could the reason be that there are many iron deposits in the region, so the "birds" left some iron residue? If so, then something whirlwind-like and/or ball-lightnings? Questions, questions, questions…

Now let's consider more established lightnings. Indeed there are trees with lightning-like damage. Here is from [Anfinogenov, and Budaeva, 1998] translated by A.O.:

"In the epicentral zone (the eastern marginal part of the Southern Swamp), continuous groves of small-sized (skinny) larches (about 150 years old) that survived the disaster are observed with peculiar spiral lesions of the trunks in 1908 (lightning, electric discharge — ?)."

Here is another account collected in 1969 [Vasil'ev et al., 1981] translated by A.O.:

"Yakochon Andrei Petrovich, Evenk, 80 years old. At the time of the disaster, he was living with his parents in Vanavara. He heard an unusual thunder, long, giving the impression of a flight. Later, the old people told him that the thunder was slow, as if something was flying. This flying body and thunder were heading north. No one knows how far it has flown. It was said that there was an extraordinary fire at the crash site; the earth and sand were burning (they were melted).<…>"

According to [Kletetschka, et al., 2019] (TE is Tunguska event):
"Paleomagnetic data revealed presence of plasma during the TE near rock surfaces".

It is interesting to note that the shape of the Kulikovskii forestfall resembles the one which is produced by a downburst. The shape of the Chuvar forestfall [Ol'khovatov, 2021] resembles the one which is produced by a burst-swath (about the burst-swath see [Fujita, 1981]). Couldn't a "collision" between them to result in the weak destruction to the west from the epicenter? Preliminary it seems not very likely as probably a strong (devastating?) turbulence is to develop in the area, but of course detailed calculations are needed for more or less reliable claims. And what about a derecho?

It is worth also to recall the following weather phenomenon which is called as the Bull's-Eye Squall. Here is from [Corliss, 1983]:

"Undoubtedly the most dangerous spout of all is the dreaded 'Bull's-Eye Squall' or 'White Squall.'
Slater Brown explains that: 'On rare occasions, when the air is too dry to produce water in a spout, though all the other conditions are present, there occurs what is known as a white or Bull's Eye Squall.... The Portugese describe a Bull's-Eye Squall as first appearing like a bright white spot at or near the zenith, in a perfectly clear sky and fine weather, and which, rapidly descending, brings with it a furious white squall or tornado. They generally occur off the west coast of Africa."

However, the Tunguska event cannot be explained by the meteorological factor alone. Indeed, according to the Evenki in the vicinity of the epicenter, the event began with seismic phenomena. Moreover, the bright lightning-like flashes (which Evenki reported) were of many times longer duration than ordinary lightnings. But that's not all. These lightning-like flashes were accompanied by powerful sound phenomena and earth tremors. Eyewitnesses reported similar sounds and earth tremors from various locations at distances of about 1000 km from the epicenter [Ol'khovatov, 2023]. Moreover in some places it happened up to noon [Ol'khovatov, 2022]. Evidences of an upsurge of tectonic activity in the region around June 30, 1908 were already presented in [Ol'khovatov, 2003; 2022]. For example, on July 5 near the settlement of Bratskoe there was a rare (for the region) earthquake M~3.5 [Ovsyuchenko, et al. 2007]. And the upsurge manifested not only in seismic phenomena over large region, but also in the following way. A. Bryukhanov, a Tomsk student, wrote from the Ust'-Kut settlement in July1908 to A.V. Voznesenskii (Director of the Irkutsk Magneto-Meteorological Observatory) (translated by A.O.):

"Only one thing is unusual and mysterious so far: a hot-salty spring with the presence of various chemicals has recently been discovered here, accidentally. Earlier, for example, last summer, it did not exist, as the peasants claim. This source was reported by me to the Irkutsk newspaper "Sibir'". "

A question about possible relations between atmospheric and endogenic processes appeared many centuries ago. Here are, for example, titles of 3 book chapters by famous F. Arago [Arago, 1855]:

"When the Atmosphere is tempestuous there are simultaneously great Perturbations in the Interior of the Earth and at the Surface or below the

Surface of Waters";

　　"The Exceptional State in which Atmospheric Storms place the Solid Part of the Globe sometimes manifests itself by Fulminating Explosions, which, without any Luminous Appearance, produce the same Effects as Thunder and Lightning properly so called";

　　"The particular State which an Atmospheric Storm communicates to the Solid and Liquid Part of the Globe by its Influence, is sometimes manifested by broad and brilliant Phenomena of Light, of which the Earth is at first the Seat, and which, after an Explosion has taken place, disappear, either by vanishing on the Spot where they were first seen, or by a more or less extensive, and more or less rapid Change of Place".

Later, interest in the study of such phenomena faded somewhat. Fortunately, in recent years, it has resumed, primarily in terms of investigating possible mechanisms of connection between earthquakes and atmospheric phenomena (discussion of possible mechanisms is beyond the scope of this paper). By the way, sometimes these atmospheric phenomena generated by tectonic processes can be rather remarkable. Here is from [Enomoto and Zheng, 1998] regarding the Kobe earthquake of 1995:

　　"A fisherman at Tonouchi on Awaji Island (Figure 1a), stated that he saw several electric streaks of bluish-white color spread out for about a second from a localized area near the Nojima fault...
　　<...>
　　We found unusual features in the fault in the Hirabayashi district where it crossed an unpaved road.
　　<...>
　　The unusual features were: 1) vegetation roots exposed on the wall of the fault were extraordinarily charred as shown by arrow A in Figure 1b; 2) sharp and blackened veins marked the weathered granite fault wall as shown by arrow B in Figure 1b; and 3) the clayey fault gouge under the part where the charred roots were found was highly lithified and showed a lamellar structure.
　　<...>
　　Furthermore, chemical analyses by an induction coupling plasma (ICP) method showed that the contents of metal elements such as Fe, Ti, Al, Mn etc. in the charred roots was about 10 times as large as that of the non-charred roots of the same vegetation."

There are also some other cases of poorly understood natural phenomena which could melt sand and metal [Ol'khovatov, 2020b].

The general planetary geophysical situation around June 30, 1908 has a number of peculiarities, which hints that in addition to the regional factor, a global factor also played a role in the phenomenon [Ol'khovatov, 2003]. It is possible that an increase in solar activity these days played a certain role also. All these are topics for future research.

A natural question arises - how often do such phenomena occur and what kind of energy can they achieve? The author of this paper does not know the answer to this question.  The study of natural phenomena in the history of our planet will help answer it.

## 4. Conclusion

The general conclusion is that the Tunguska event was a very complex phenomenon. A reader is welcome to make his own conclusions regarding the accounts and various manifestations of Tunguska. Anyway any interpretation of the Tunguska event should explain the accounts and the manifestations.

**ACKNOWLEDGEMENTS**

The author wants to thank the many people who helped him to work on this paper, and special gratitude to his mother  - Ol'khovatova Olga Leonidovna (unfortunately she didn't live long enough to see this paper published...), without her moral and other diverse support this paper would hardly have been written.